\RequirePackage{fix-cm}
\documentclass[twocolumn,epjc3]{svjour3m}
\usepackage{pdfsync}
\usepackage{epsfig,epsf}
\usepackage{amsmath}
\usepackage{amsfonts}
\usepackage{amssymb}
\usepackage{dsfont}
\usepackage{slashed}
\usepackage{cancel}
\usepackage{epstopdf}
\smartqed  
\DeclareMathOperator{\tr}{tr}

\newcommand \vev [1] {\langle{#1}\rangle}
\newcommand \VEV [1] {\left\langle{#1}\right\rangle}

\def\II{\hbox{{1}\kern-.25em\hbox{l}}}

\RequirePackage{graphicx}
\RequirePackage[numbers,sort&compress]{natbib}
\journalname{DESY 15-038}

\begin{document}

\title{Conformal constraints for anomalous dimensions of leading twist operators
}


\author{A.~N.~Manashov\thanksref{e1,addr1,addr2,addr3}
        \and
        M.~Strohmaier\thanksref{e2,addr2} 
}

\thankstext{e1}{e-mail:  alexander.manashov@desy.de}
\thankstext{e2}{e-mail: matthias.strohmaier@ur.de}


\institute{Institut f\"ur Theoretische Physik Universit\"at Hamburg, Luruper Chaussee 149,
            D-22761 Hamburg, Germany \label{addr1}
            \and
Institut f\"ur Theoretische Physik, University of Regensburg, D-93040 Regensburg, Germany  \label{addr2}
\and
         Department of Theoretical Physics,  St.-Petersburg State University,
          199034 St.-Petersburg, Russia \label{addr3}
}

\date{Received: date / Accepted: date}

\maketitle

\begin{abstract}
Leading-twist operators have a remarkable property that their divergence vanishes in a free theory. Recently it
was suggested that this property can be used for an alternative technique to calculate anomalous dimensions of
leading-twist operators and allows one to gain one order in perturbation theory so that, i.e., two-loop anomalous
dimensions can be calculated from one-loop Feynman diagrams, etc. In this work we study feasibility of this
program on a toy-model example of the $\varphi^3$ theory in six dimensions. Our conclusion is that this approach
is valid, although it does not seem to present considerable technical simplifications as compared to the standard
technique. It does provide one, however, with a very nontrivial check of the calculation as the structure of the
contributions is very different.

\end{abstract}

\section{Introduction}

Calculation of anomalous dimensions of composite operators belongs to
the standard tasks of any quantum field theory calculation.
For example, in quantum chromodynamics, anomalous dimensions of
leading twist two operators govern the scaling behavior of quark and gluon
distributions in hadrons and have to be known with high precision.
Nowadays the anomalous dimensions are known at three-loops,
see Refs.~\cite{Moch:2004pa,Vogt:2004mw,Moch:2014sna} and references therein. Beyond the two-loop approximation  such calculations
are feasible only with the help of the advanced  methods of computer algebra.
Since the calculations are fully automated finding errors becomes highly nontrivial task and any approach which can provide a  check of  the final results is very helpful.

Given that the theory depends not only on the coupling constant but also some other parameters such as the dimension of
an internal symmetry group, one can organize  expansion over these parameters.
The best known example of this kind is the $1/N$ expansion, see Ref.~\cite{Moshe:2003xn} for a review.
Agreement between the results  obtained in the perturbative and $1/N$ expansions serves as  a
powerful test  for the  validity of calculations. However, the  calculations in the $1/N$ expansion are much harder than perturbative
calculations: only two RG functions -- indices of the basic fields in the nonlinear $\sigma-$model and Gross -- Neveu model -- are available at
$1/N^3$ order~\cite{Vasiliev:1982dc,Vasiliev:1992wr,Gracey:1993kc}. In QCD the calculations rarely go beyond the
leading order in $1/N_f$ (where $N_f$ is the number of flavors). At leading order in $1/N_f$ the anomalous
dimensions of twist two operators  were calculated in Refs.~\cite{Gracey:1994nn,Bennett:1997ch} but extension of
these results to the next order is hardly possible.

A new approach for calculating the anomalous dimensions of
leading twist operators was  proposed in Refs.~\cite{Anselmi:1998ms,Belitsky:2007jp}.
It is still a perturbative approach, however, the contributing diagrams are completely different from those in the
standard technique.
The approach  is based on a remarkable  property of leading twist  operators: namely,  a
divergence of such an  operator, $\mathcal{O}_{\mu_1,\ldots \mu_j}$,  vanishes in a free
theory~\cite{Ferrara:1972xq}
\begin{align}\label{FGPG}
\partial^{\mu_1}\mathcal{O}_{\mu_1,\ldots \mu_j}(x)=0\,.
\end{align}
In the interacting theory the r.h.s. of Eq.~(\ref{FGPG}) is non-zero but is proportional to the coupling constant.
This identity allows to extract the $\ell-$loop contribution to the anomalous dimension of the operator
$\mathcal{O}_{\mu_1,\ldots \mu_j}$ from    $\ell-1$ loop diagrams  only. In particular the one loop anomalous
dimensions of leading twist operators do not require calculation of loop integrals at
all~\cite{Belitsky:2007jp,Braun:2011dg}.

The method developed in Ref.~\cite{Belitsky:2007jp} is adjusted to local operators and relies heavily upon the so-called "conformal scheme"
renormalization~\cite{Mueller:1991gd,Mueller:1993hg,Mueller:1997ak}.  In our opinion it is more convenient
 to stay within the standard
$\overline{\textrm{MS}}$ scheme and  the formalism   non-local (light-ray) operators technique. This technique proves to be  more effective
and flexible as we will demonstrate on the example of calculation  of two -- loop anomalous dimensions
in the  $su(n)$ symmetric $\varphi^3$ model~\cite{Braun:2013tva}.

The paper is organized as follows. In section~\ref{sect:background} we introduce the model and fix  notations.
In section~\ref{sect:light-ray} we recall the light-ray operator technique. Section~\ref{sect:divergence} is devoted
to calculation of the divergence of conformal operator. The details of calculation of two -- loop correlators are presented in
section~\ref{sect:correlators}.  Our conclusions are  in section~\ref{sect:summary}.
In  Appendices  we explain some technical issues and details of the derivation.

\section{Generalities}\label{sect:background}
The $su(n)$ symmetric $\varphi^3$ model is a scalar  field  theory in $d=6-2\epsilon\equiv 2\mu$ dimension with an action
\begin{align}
S(\varphi)
=&\int d^dx\left[\frac12(\partial\varphi^a)^2+\frac{1}6gM^\epsilon d^{abc}\varphi^a\varphi^b\varphi^c\right]\,,
\end{align}
where $a=1,\ldots, n^{2}-1$,
\begin{align}
d^{abc}=2\tr t^a\{t^b t^c\}\,,
\end{align}
$t^{a}$ are the generators of the $su(n)$ algebra
normalized in the conventional way, $\tr t^a t^b =1/2$. The theory is multiplicatively renormalizable
\begin{align}
S_R(\varphi,g)=S(\varphi_0,g_0)\,,
\end{align}
where
$$
\varphi_0=Z_\varphi \,\varphi\qquad \text{and} \qquad g_0=M^\epsilon Z_g  g\,.
$$
Two--loop expressions for
the renormalization constants $Z_1=Z_\varphi^2$ and $Z_3= Z_g Z_\varphi^3$ can be found in
Ref.~\cite{Braun:2013tva}. The $\beta-$function of the charge $u=g^2/(4\pi)^3$ and the field anomalous dimension
$\gamma_\varphi$
are
\begin{align}
\beta(u)&=-2\epsilon u-u^2\frac{n^2-20}{2n}
+\mathcal{O}(u^3)\,,
\notag\\
\gamma_\varphi(u)& = u\frac{n^2-4}{12 n}\left(1+u\frac{n^2-100}{36 n}\right)+\mathcal{O}(u^3)\,.
\label{rgf-3}
\end{align}
At the critical point $u_*$, $\beta(u_*)=0$,
\begin{align}\label{ust3}
u_*={4n\epsilon}/(20-n^2)+ O(\epsilon^2)
\end{align}
theory enjoys the scale and conformal invariance.~\footnote{Formally, a nontrivial critical point for $d<6$ only exists for $n=3,4$.
However, staying within perturbation theory one can consider $n$ as a continuous parameter. In this sense
all further results  hold for arbitrary $n$.}

It is well known that in a conformal theory the form of two--point correlation function of conformal operators is  fixed  up to normalization.
In particular, the correlator of the conformal  traceless symmetric operators  has the form
\begin{align}\label{OO}
\VEV{\mathcal{O}^{(n)}_{j}(x)\mathcal{O}^{(\bar n)}_{j'}(y)}=\delta_{jj'}\delta_{\Delta_j\Delta_{j'}}\frac{C_jI_{n,\bar n}^j(x-y)}{((x-y)^2)^{\Delta_j}}\,.
\end{align}
Here $j$ $(j')$ is the spin of the operator. The vectors $n$ and $\bar n$ are two light--like vectors, $n^2=\bar
n^2=0,$ and
$\mathcal{O}^{(n)}_j(x)~(\mathcal{O}^{(\bar n)}_{j'}(y))$ is a contraction of an
operator with  the vector $n~(\bar n)$, for instance
\begin{align}
\mathcal{O}^{(n)}_j(x)=n^{\mu_1}\ldots n^{\mu_j} \mathcal{O}_{\mu_1\ldots\mu_j}(x)\,.
\end{align}
$\Delta_j$ and $\Delta_{j'}$
are the scaling dimensions of the operators, $C_j$ is a normalization constant and
\begin{align}
I_{n,\bar n}(x)=(n\bar n)-\frac{2(nx)(\bar n x)}{x^2}\,.
\end{align}
The scaling dimension of the operator is given by the sum of  canonical and anomalous dimensions at a
critical point, $\Delta_j=\Delta_j^{(0)}+\gamma_j$, where $\gamma_j\equiv \gamma_j(u_*)$. For the leading twist operators of spin $j$,
$\Delta_j^{(0)}=2\mu-2+j$.

The anomalous dimension of an operator in MS scheme, $\gamma_j(u)$,  is a function of a coupling constant only.
It can be restored from its critical value, $\gamma_j=\gamma_j(u_*)$, provided that
the latter is known as a function of $\epsilon$.

The correlation function for the divergence of the conformal operator
\begin{align}\label{divergence}
\partial \mathcal{O}^{(n)}_j(x)\equiv j\,n^{\mu_2}\ldots n^{\mu_j}\,\partial^{\mu_1}\mathcal{O}_{\mu_1\mu_2\ldots\mu_j}(x)
\end{align}
can be obtained from Eq.~(\ref{OO}). If $x$ is chosen in the transverse plane, $(xn)=(x\bar n)=0$,  the ratio of
the two correlation functions
\begin{align}\label{OO/OO}
\mathcal{T}_j(u_*)&=x^2(n\bar n)
\frac{\VEV{\partial\mathcal{O}_j^{(n)}(x)\partial\mathcal{O}^{(\bar n)}_j(0)}}{\VEV{\mathcal{O}_j^{(n)}(x)\mathcal{O}_j^{(\bar n)}(0)}}
\end{align}
is a function of the anomalous dimension $\gamma_j$ only
\begin{align}\label{t-gamma-eq}
\mathcal{T}_j&=
2j\gamma_j
\left[\frac{(2\mu-3+j)(\mu-1+j)}{\mu-2+j}+2\gamma_j\right].
\end{align}
This ratio, in full agreement with the result of Ref.~\cite{Ferrara:1972xq},  vanishes provided that $\gamma_j=0$.

The perturbation series for $\gamma_j$ and $\mathcal{T}_j$
\begin{align}\label{pert-exp}
\gamma_j&=u_* \,\gamma_j^{(1)}+u_*^2 \,\gamma_j^{(2)}+\ldots\,,
\notag\\[2mm]
\mathcal{T}_j&=
\varkappa_j\left(u_*\,{T}_j^{(1)}+u_*^2 \, {T}_j^{(2)}+\ldots\right)\,,
\end{align}
are related to each other. For later convenience we chose the normalization factor $\varkappa_j$ as follows
\begin{align}
\varkappa_j=\frac{2j(j+2)(j+3)}{j+1}\,.
\end{align}
Substituting the series \eqref{pert-exp} into Eq.~\eqref{t-gamma-eq}
one finds  the following relations between the  expansion coefficients:
\begin{align}
\gamma_j^{(1)}&={T}_j^{(1)}\,,
\notag\\
\gamma_j^{(2)}&={T}_j^{(2)}-\frac{2(j+1)}{(j+2)(j+3)} (T_j^{(1)})^2
\notag\\
&\quad
+\frac{2j^2+5j+1}{(j+1)(j+2)(j+3)} \epsilon^{(1)}\,T_j^{(1)}\,,
\end{align}
and so on. Here $\epsilon^{(1)}=(n^2-20)/4n$.

Since the divergence of the conformal operator is proportional to the coupling constant, $\partial\mathcal{O}_j^{(n)}\sim
O(g)$, the ratio $\mathcal{T}_j$ contains a "kinematical" factor $u\sim g^2$.
Thus, in order to determine  $\mathcal{T}_j$ and, hence, the anomalous dimension $\gamma_j$,
 with $O(u^\ell)$ accuracy the corresponding correlation functions
have to be calculated at one order in  $u$ less.
 In Ref.~\cite{Belitsky:2007jp} one loop anomalous dimensions were reproduced by this method in $\varphi^3$ and $N=4$
SUSY models. Going beyond the leading order requires an effective technique for calculation of the two--point
correlation functions
otherwise one gains nothing in comparison with the standard approach.

\section{Light-ray vs local operators}\label{sect:light-ray}
The first task is to find a convenient description for local operators.  As we will argue
the light-ray operator technique~\cite{Balitsky:1987bk} is a most suitable one.
The light-ray operator~\footnote{In order to make presentation more transparent we will consider the $su(n)$ scalar
operator. The operators of other symmetry properties can be easily included into consideration, see
Sect.\ref{sect:results}.
 }
\begin{align}\label{dO}
[\mathcal{O}(x;z_1,z_2)]&=[\varphi^{a}(x+z_1 n)\varphi^{a}(x+z_2 n)]
\end{align}
is defined as the generating function for the renormalized local operators
\begin{align}\label{defO}
[\mathcal{O}(x;z_1,z_2)]
&\equiv \sum_{k,m} z_1^k z_2^m [\mathcal{O}_{km}](x)
\notag\\
&=\sum_{k,m,k',m'} z_1^k z_2^m\, Z_{km}^{k'm'}\,\mathcal{O}_{k'm'}(x)\,.
\end{align}
Here $[\mathcal{O}_{km}]$ is the renormalized (in MS scheme) local monomial  $\mathcal{O}_{km}=
\partial_+^k\varphi^a(x)\partial_+^m\varphi^a(x)/{k!m!}$
 and   $\partial_+ =(n \partial_x)$.
The sum in Eq.~(\ref{defO}) can be replaced by action of some integral operator  on the
bare  operator%
\begin{align}\label{light-ray}
[\mathcal{O}(x;z)]=Z\mathcal{O}(x;z)\,.
\end{align}
Here we introduced a shorthand notation $z=\{z_1,z_2\}$.
The integral operator $Z$ can be written in the form~\cite{Braun:2013tva}
\begin{align}
Zf(z)=\int d\alpha d\beta \, Z(\alpha,\beta)\, f(z_{12}^\alpha,z_{21}^\beta)\,.
\end{align}
The renormalization kernel $Z(\alpha,\beta)$ is given by  a series in $1/\epsilon$ and the coupling~$u$.
The light-ray operator~(\ref{defO}) satisfies the RG equation
\begin{align}
\Big(M\partial_M +\beta(u)\partial_u+ \mathbb{H}(u)\Big)[\mathcal{O}(x;z)]=0\,,
\end{align}
where the  evolution kernel $\mathbb{H}$ is an integral operator
\begin{align}
\mathbb{H}(u)=-\left(M\frac{d}{dM} Z\right)\,Z^{-1}+2\gamma_\varphi\,
\end{align}
which encodes  all information on  the anomalous dimension matrices for local operators.

At the critical point $u=u_*$ local operators can be classified according to the representations of the conformal group.
An operator with the lowest scaling dimension in the representation is called a conformal operator. The leading
twist operator~\footnote{Note, that the operator $\mathcal{O}_j$ vanishes identically  for odd $j$.}, $\mathcal{O}_j$, is uniquely determined by its scaling dimension $\Delta_j$. The expansion
of the
light-ray operator~(\ref{defO}) over conformal operators and their descendants reads~\cite{Braun:2013tva}
\begin{align}\label{expansion}
[\mathcal{O}(x;z)]=\sum_{jk}\Psi_{jk}(z_1,z_2) \,\partial_+^k \mathcal{O}_j(x)\,,
\end{align}
where the coefficients $\Psi_{jk}(z_1,z_2)$ are homogeneous polynomials of degree $j+k$ in $z_1,z_2$. These
polynomials are eigenfunctions of the evolution kernel $\mathbb{H}(u_*)$
\begin{align}\label{eigenH}
\mathbb{H}(u_*)\Psi_{jk}(z)=\gamma_j\,\Psi_{jk}(z)\,,
\end{align}
with the corresponding eigenvalues being the anomalous dimensions, $\gamma_j=\gamma_j(u_*)$.
Since  the theory enjoys conformal invariance at the critical point $u=u_*$ the evolution kernel 
commutes with three generators
of the collinear subgroup of  conformal group,~\footnote{Other generators acts  on the operators in
question trivially.}
\begin{align}\label{SH}
[S_{\pm,0},\mathbb{H}(u_*)]=0.
\end{align}
 The generators, however, deviate from their canonical form  (see Appendix~\ref{app:A})
\begin{align}
S_-^{(0)}&=-\partial_{z_1} - \partial_{z_2}\,,
\notag\\
S_0^{(0)}&=z_1\partial_{z_1}+z_2\partial_{z_2}+2\,,
\notag\\
S_+^{(0)}&=z_1^2\partial_{z_1}+z_2^2\partial_{z_2}+2(z_1+z_2)\,
\end{align}
due to quantum corrections,
\begin{align}\label{fullS}
S_{\pm,0}=S_{\pm,0}^{(0)}+\Delta S_{\pm,0}.
\end{align}
 Two of the generators are known to all orders,
\begin{align}\label{delta-S}
\Delta S_-=0\,, &&\Delta S_0=-\epsilon+\frac12 \mathbb{H}(u_*)\,,
\end{align}
while  corrections to the generator of special conformal transformations can be calculated order by order in
perturbation theory~\cite{Braun:2013tva}. The leading correction is
\begin{align}
\Delta S_+=(z_1+z_2)\left(-\epsilon+\frac12 \mathbb{H}(u_*)\right)+O(\epsilon^2)\,.
\end{align}
It follows from Eqs.~(\ref{eigenH}),~(\ref{SH}) that the operators  $S_\pm$ act as raising (lowering) operators on the
set of  eigenfunctions $\Psi_{jk}$,
%
\begin{align}\label{SPM}
S_\pm\Psi_{jk}\sim \Psi_{jk\pm1}\,.
\end{align}
In turn $S_0$ counts conformal spin of the operator $\mathcal{O}_{jk}$,
\begin{align}
S_0\Psi_{jk}=j_{jk}\Psi_{jk}=\frac12\left(\Delta_{jk}+S_{jk}\right)\Psi_{jk}\,,
\end{align}
where $\Delta_{jk}=\Delta_j+k$ and $S_{jk}$ are the scaling dimension and spin of the operator, respectively.
One derives immediately from~(\ref{SPM}) that the polynomial $\Psi_{jk=0}$ accompanying the conformal operator $\mathcal{O}_j$
in the expansion~(\ref{expansion})
is a
simple power, $\Psi_{jk=0}(z_1,z_2)\sim(z_1-z_2)^j$ and all other eigenfunctions have the form $\Psi_{jk}(z_1,z_2)\sim S_+^k (z_1-z_2)^j$.

\subsection{Scalar product}
Eq.~(\ref{eigenH}) can be considered as a standard quantum mechanical problem for the Hamiltonian
$\mathbb{H}$. In order to make this analogy complete one needs to introduce a scalar product on the space of the
eigenfunctions. Clearly, such a scalar product has to be adjusted to the symmetries of the problem. At leading order
the answer is given by the standard $sl(2)$ invariant scalar product~\cite{GelGraVil66,Vilenkin}
\begin{align}\label{sc0}
(\psi,\phi)_{0}=\frac1{\pi^2}\iint\limits_{|z_k|<1} d^2z_1 d^2z_2 \,(\psi(z_1,z_2))^\dagger \,\phi(z_1,z_2)\,.
\end{align}
The integration goes over the unit disks $|z_{k}|<1$, $k=1,2$.
The generator $S_0^{(0)}$ is a self-adjoint operator with respect to this scalar product and
$(S_+^{(0)})^\dagger=-S_-^{(0)}$. We want to find a deformation of the scalar product~(\ref{sc0})  that
keeps these   relations for the complete generators, $S_0=S_0^\dagger$ and $ S_+^\dagger=-S_-$. Let us look for the
solution in the form
\begin{align}\label{sc}
(\psi,\phi)_{\varpi}=(\psi, \varpi\phi)_0\,, &&\varpi=\II+ u_* \varpi^{(1)}+\ldots\,,
\end{align}
where $\varpi^{(1)}$ is a self-adjoint operator with respect to the scalar product~(\ref{sc0}). The conjugation
conditions for the generators imply
\begin{align}\label{SDelta}
\Delta S_0^{(1)}-(\Delta S_0^{(1)})^\dagger=[S_0^{(0)},\varpi^{(1)}]\,,
\notag\\
\Delta S_+^{(1)}=[S_+^{(0)},\varpi^{(1)}]\,.
\end{align}
The one loop corrections to the generators involve the kernel
 $\mathcal{H}^{(1)}$ $\big(\mathbb{H}(u)=\sum_k u^k\mathcal{ H}^{(k)}\big)$ which is given by the following
expression
\begin{align}\label{H-one-loop}
\mathcal{ H}^{(1)}=2(\gamma^{(1)}_\varphi-\lambda_s \mathcal{H}^+)\,.
\end{align}
Here $\lambda_s$ is a color factor,
$\lambda_s=(n^2-4)/n$ and
\begin{align}\label{H+}
\mathcal{H}^+\psi({z})=\int_0^1d\alpha\int_0^{\bar\alpha}d\beta\,\psi(z_{12}^\alpha,z_{21}^\beta)\,.
\end{align}
The operator $\varpi^{(1)}$ is completely determined by Eqs.~(\ref{SDelta}).  The explicit expression for $\varpi^{(1)}$
and details of the derivation can be found in~\ref{app:B}~\footnote{It turns out that  the
corrections due to $\varpi^{(1)}$ cancel at $O(\epsilon^2)$ order in the ratio of the correlators~(\ref{OO/OO}).
So that we do not need this explicit expression for the present purposes.}.

Since the eigenfunctions $\Psi_{jk}$ are mutually orthogonal w.r.t. the scalar product~(\ref{sc}), one can represent
the conformal operator as the scalar product of the coefficient function with the light-ray operator
\begin{align}\label{scalar-O}
\mathcal{O}_j(x)=(z_{12}^j, [\mathcal{O}(x,z)])_{\varpi}\,.
\end{align}
This representation for the conformal operator is the most convenient one for further analysis.
We demonstrate it on the following example. The conformal operator is usually defined as an operator which
vanishes under special conformal transformations, $K_{\bar n}=K\cdot\bar n$, $\delta_{K_{\bar n}}\mathcal{O}_j(0)=0$.
This property becomes transparent in the representation~(\ref{scalar-O}) if one takes into account that
$$
\delta_{K_{\bar n}}[\mathcal{O}(z)]=2(n\bar n) S_+[\mathcal{O}(z)],
$$
(we put here $[\mathcal{O}(z)]=[\mathcal{O}(x=0;z)]$) and use that the generators $S_+$ and $S_-$ are conjugate to each
other w.r.t. the scalar product~(\ref{sc}), $S_+^\dagger=-S_-$.

\section{Divergence of conformal operator}\label{sect:divergence}
In order to construct the divergence of the conformal operator $\partial \mathcal{O}_j^{(n)}$,   Eq.~(\ref{divergence}),
we calculate first the divergence of the light-ray operator~(\ref{light-ray})
\begin{align}\label{DO}
[\partial \mathcal{O}(x;z)]\equiv \frac\partial{\partial x^\mu}\frac\partial{\partial n_\mu}[\mathcal{O}(x;z)]\,.
\end{align}
Taking the $n-$derivative one cannot, however, keep $n^2=0$ any longer and has to take into account terms linear in
$n^2$ which account for the trace subtraction in  $[\mathcal{O}(x;z)]$. Taking the corresponding modification into
account, see e.g.  Refs.~\cite{Belitsky:2007jp,Balitsky:1987bk,Bargmann:1977gy}, one gets for~(\ref{DO})
\begin{align}
[\partial \mathcal{O}(x;z)]&= \frac\partial{\partial x^\mu}\nabla_\mu Z\varphi^{a}(x+z_1 n)\varphi^{a}(x+z_2 n)
\end{align}
where $\nabla^\mu$ is a differential operator
\begin{align}
\nabla_\mu=\frac{\partial}{\partial n^\mu}-\frac12(\mu-1+n\cdot\partial_n)^{-1} n^\mu\frac{\partial^2}{\partial n^2}\,.
\end{align}
which  commutes with the renormalization factor $Z$ and acts on the fields directly.
After a simple algebra one gets
\begin{align}\label{div-light-ray}
[\partial \mathcal{O}(x;z)]&=\frac1{2} (S_0^{(\epsilon)}-1)^{-1}Z\Biggl\{ S_+^{(\epsilon)} \,\partial_x^2\, \mathcal{O}(x;z)
\notag\\
&\quad-L^{(\epsilon)}_{21}\, \partial^2\varphi^a(x+z_1n)\varphi^a(x+z_2n)
\notag\\
&\quad-L_{12}^{(\epsilon)}\,\varphi^a(x+z_1n)\partial^2\varphi^a(x+z_2n)\Biggr\}\,,
\end{align}
where $S_0^{(\epsilon)}=S_0^{(0)}-\epsilon$,  $S_+^{(\epsilon)}=S_+^{(0)}-\epsilon(z_1+z_2)$ and
\begin{align}
L_{21}^{(\epsilon)}=\partial_{z_2} z_{21}^{2}-\epsilon z_{21}\,, &&L_{12}^{(\epsilon)}&=\partial_{z_1} z_{12}^{2}-\epsilon z_{12}\,.
\end{align}
Using equations of motion (EOM) one can replace in this expression
\begin{align}
\partial^2\varphi^a(x)\mapsto\frac12 g M^\epsilon Z_3 Z_1^{-1}d^{abc}\varphi^b(x)\varphi^c(x)\,.
\end{align}
We want to stress here that Eq.~(\ref{div-light-ray}) holds for arbitrary  coupling $u$ but not only at the
critical value. Since the l.h.s. of Eq.~(\ref{div-light-ray}) is a finite (renormalized) operator the r.h.s. can be
expressed in terms of  renormalized operators with finite coefficients. These operators can be chosen as:  the two-particle
operator $\mathcal{O}_1=\partial^{2}\mathcal{O}(x;z)$ and three particle operator
\begin{align}
\mathcal{O}_2&=\mathcal{O}^{(d)}(x;w)
\notag\\
&=gd^{abc}\varphi^a(x+w_1n)\varphi^b(x+w_2n)\varphi^c(x+w_3n)\,,
\end{align}
where $w=\{w_1,w_2,w_3\}$. The operators $\mathcal{O}_1$ and $\mathcal{O}_2$  mix under renormalization. The mixing matrix (integral operator
acting on  fields variables) has an lower
triangular form
\begin{align}
[\mathcal{O}_k]=Z_{km} \mathcal{O}_m\,.
\end{align}
Here $Z_{11}=Z$ is the renormalization constant of the light-ray  operator, Eq.~(\ref{light-ray}),
$Z_{12}=0$, $Z_{21}=O(u^2)$ and  the element $Z_{22}$  is given, at the one loop order, by the sum of two - particles kernels
\begin{align}\label{Zfactors}
Z_{11}&=1-\frac{u}{\epsilon}\,\lambda_s\,\mathcal{H}_{12}^{+}+O(u^2)\,,
\notag\\
Z_{22}&=\II+\frac{u}{2\epsilon}\sum\nolimits_{i<k}\left(\lambda_s\mathcal{H}_{ik}^{d}-
\lambda_d\,\mathcal{H}_{ik}^{+}\right)+
O(u^2)\,,
\end{align}
where $\lambda_s,\lambda_d$ are  color factors
\begin{align}
\lambda_s=(n^2-4)/n\,, &&\lambda_d=(n^2-12)/n\,.
\end{align}
The kernel $\mathcal{H}_{ik}^+$ is defined by Eq.~(\ref{H+})
and  $\mathcal{H}_{ik}^d$ has the form
\begin{align}
\mathcal{H}_{12}^{d}f(z_1,z_2)=\int_0^1d\alpha\, \alpha\bar\alpha \, f(z_{12}^\alpha,z_{12}^\alpha)\,.
\end{align}
The subscripts $ik$ show the arguments the  kernel
acts on.

Using these results we can rewrite~(\ref{div-light-ray}) as follows
\begin{align}\label{Div-light-ray}
[\partial \mathcal{O}(x;z)]&=\frac1{2} (S_0^{(\epsilon)}-1)^{-1}\sum\nolimits_{k=1,2} A_k [\mathcal{O}_k(x;z)]\,.
\end{align}
The operators $A_k$ have the following form:
\begin{align}
A_1&=Z_{11} S_+^{(\epsilon)} Z_{11}^{-1}-M^{-\epsilon} A_2 Z_{21} Z_{11}^{-1}\,,
\notag\\
A_2&=-\frac12M^\epsilon Z_3 Z_1^{-1}Z_{11}\Big(L_{12}^{(\epsilon)}S_2+L_{21}^{(\epsilon)} S_1\Big)Z_{22}^{-1}\,,
\end{align}
where the operators $S_1,S_2$ map functions of three variables to  functions of two variables
\begin{align}
[S_1 f](z_1,z_2)&=f(z_1,z_1,z_2)\,, \notag\\
[S_2 f](z_1,z_2)&=f(z_1,z_2,z_2)\,.
\end{align}
At one loop the operators $A_k$ take the form
\begin{align}\label{A12}
A_1&=S_+(u) + {u}\Big(\lambda_s \mathcal{H}^+-\gamma_\varphi^{(1)}\Big)(z_1+z_2)+O(u^2)\,,
\notag\\
A_2&=-\frac12 M^\epsilon\Big( L_{12}S_2+L_{21} S_1
+\Big(\epsilon-u\lambda_s \mathcal{H}^+ \Big)z_{12}S_{12}
\notag\\
&\quad-u z_{12} S_{12} \Big(\lambda_s \mathcal{H}_{13}^d-\lambda_d \mathcal{H}^+_{13}\Big)+O(u^2)\Big)\,.
\end{align}
Here $S_+(u)$ is given by the expressions~(\ref{fullS}), (\ref{delta-S}) for arbitrary $u$, $u_*\to u$,
%
$$
S_{12}=S_1-S_2\,, \qquad L_{km}=L_{km}^{(\epsilon\mapsto 0)}
$$
%
and deriving~(\ref{A12})  we made use of the symmetry of the three particle operator $\mathcal{Q}^{(d)}(x;w_1,w_2,w_3)$
under permutation of $w$ variables.

Let us stress again that the operators $A_k$ do not contain singular terms for \emph{arbitrary} $u$.
Using one loop expressions for $Z$ factors, Eq.~(\ref{Zfactors}),
it can
be checked that all pole terms cancel at  order $O(u)$. In particular
\begin{multline}\label{LZ=ZL}
Z_3 Z_1^{-1}Z_{11}\Big(L_{12}S_2+L_{21} S_1\Big)Z_{22}^{-1}=\\=
L_{12}S_2+L_{21} S_1+O(u^2)\,.
\end{multline}

Starting from the representation~(\ref{scalar-O}) for  the conformal operator, we get for its divergence
\begin{align}
\partial\mathcal{O}_j(x)=(z_{12}^j,[\partial \mathcal{O}(x;z)])_{\varpi}\,.
\end{align}
Making use of Eqs.~(\ref{Div-light-ray})~--~(\ref{A12}) one finds that the divergence $\partial\mathcal{O}_j(x)$ is given
 by the sum of two--particle and
three--particle (renormalized) operators
\begin{align}\label{OK23}
\partial\mathcal{O}_j(x)=\frac{1}{2(j+1-\epsilon)}\Big(\mathcal{R}^{(2)}_j(x)+\mathcal{R}^{(3)}_j(x)\Big)\,.
\end{align}
The prefactor on the r.h.s of Eq.~(\ref{OK23}) is the eigenvalue of the operator $(S_0^{(\epsilon)}-1)^{-1}$
on the function $z_{12}^j$.  The two--particle term $\mathcal{R}_j^{(2)}$ has the form
\begin{align}\label{K2}
\mathcal{R}^{(2)}_j(x)&= {\gamma_j}\,\bigl(z_{12}^j,\, (z_1+z_2) [\mathcal{O}_1(x;z)]\bigr)+O(u_*^2).
\end{align}
We recall that $\gamma_j=\gamma_j(u_*)$ and  $\mathcal{O}_1(x;z)=\partial^2 \mathcal{O}(x;z)$.
Let us note that  the term $\sim S_+$ in the expression for $A_1$,
Eq.~(\ref{A12}), vanishes inside the scalar product since $(z_{12}^j, S_+\ldots)=-(S_-z_{12}^j,\ldots)=0$.
In  turn, the expression for the three--particle contribution can be written as follows
\begin{align}\label{K3}
\mathcal{R}^{(3)}_j(x)=\mathcal{R}^{(3, 0)}_j(x)+\mathcal{R}^{(3,1)}_j(x)+O(\epsilon^2)\,,
\end{align}
where
\begin{align}\label{R23}
\mathcal{R}^{(3, 0)}_j(x)&=-{M^\epsilon}(z_{12}^j,\,L_{21}S_1 [\mathcal{O}_2(x;z)]\bigr)_{\varpi}\,,
\notag\\
\mathcal{R}^{(3,1)}_j(x)&=-{M^\epsilon}(z_{12}^j,z_{12} S_1 X_j[\mathcal{O}_2(x;z)]\bigr)_{\varpi}
\end{align}
%
and the operator $X_j$ has the form
\begin{align}
X_j=\epsilon-\gamma_\varphi+\gamma_j/2 -u_*\big(\lambda_s \mathcal{H}_{13}^d-\lambda_d \mathcal{H}^+_{13}\big)\,.
\end{align}
%
%
This expression follows immediately from~(\ref{A12}) if one takes into
account that  spin $j$ is even
and $A_2$ is symmetric under interchange $z_1\leftrightarrow z_2$.

It is clear from~(\ref{K2}) that the expansion of two particle term $\mathcal{R}^{(2)}_j$ over conformal
operators does not contain the operator of  spin $j$,
\begin{align}
\mathcal{R}_j^{(2)}\sim \partial^2\left(\sum_{m=0}^{j-2} c_m  \partial^{j-m-2}_+
\mathcal{O}_m(x)\right)\,.
\end{align}
Taking into account that $\vev{\mathcal{O}_j(x)\mathcal{O}_k(0)}=0$ for $k<j$
one derives  that $\vev{\mathcal{R}_j^{(2)}(x)\, \partial
\mathcal{O}_j(0)}=0$. This, in virtue of Eq.~\eqref{OK23}, results in the following relation for  the correlators
\begin{align}
\vev{\mathcal{R}_j^{(2)}(x)\mathcal{R}_j^{(2)}(0)}=-\vev{\mathcal{R}_j^{(2)}(x)\mathcal{R}_j^{(3)}(0)}\,.
\end{align}
It can be shown that in the correlator $\vev{ \partial\mathcal{O}_j(x) \partial\mathcal{O}_j(0)}$  one can replace~(\ref{OK23})
by a simpler expression
\begin{align}
\partial\mathcal{O}_j(x)=\frac{1}{2(j+1)}\Big(\mathcal{R}^{(2)}_j(x)+\mathcal{R}^{(3,0)}_j(x)\Big)\,. 
\end{align}
The omitted terms
$$
\Delta \mathcal{X}=\frac\epsilon{j+1}\left(\mathcal{R}^{(3,1)}_j(x)+\frac1{j+1}\mathcal{R}^{(3,0)}_j(x)\right)+O(\epsilon^2)
$$
give rise to the correction of order $O(\epsilon^3)$. In order to verify this  it is sufficient to
notice that $\Delta \mathcal{X}$
can be rewritten in the form  $(z_{12}^j, F\, S_- z_{12} [\mathcal{O}_2(x;z)]\bigr)_{\varpi} $,
where $F$ is some operator whose explicit expression is not relevant.
Inside the correlator the generator $S_-$ acts, finally, on the function $z_{12}^j$
nullifying it.

Thus in order to find the anomalous dimension $\gamma_j$ at order $O(\epsilon^2)$ one has to calculate the three correlators
$$
\vev{\mathcal{O}_j\mathcal{O}_j}, \qquad \vev{\mathcal{R}_j^{(2)}\mathcal{R}_j^{(3,0)}}, \qquad \vev{\mathcal{R}_j^{(3,0)}\mathcal{R}_j^{(3,0)}}
$$%
at one loop order.  We will do it in the next section.

Finally, we note that for $j=2$ the r.h.s. of
Eq.~(\ref{OK23}) has to vanish identically since the operator $\mathcal{O}_{\mu\nu}$ is, up to EOM terms, the energy-momentum tensor.
The two particle term $\mathcal{R}_{j}^{(2)}$ is proportional to the anomalous dimension $\gamma_j$ and therefore
vanishes for $j=2$. In order to check it for $\mathcal{R}_j^{(3)}$, it is sufficient to take into account that only the linear term in
the expansion of three particle operator
\begin{align*}
\mathcal{O}_2(x; w)&\sim (w_1+w_2+w_3)\cdot
d^{abc}\partial_+\varphi^a(x)\varphi^b(x)\varphi^c(x)
\\
&= \left(S_+^{(1,1,1)}\cdot 1\right)\,\, d^{abc}\partial_+\varphi^a(x)\varphi^b(x)\varphi^c(x)
\end{align*}
contributes to (\ref{K3}) for $j=2$. After simple algebra one finds that $\mathcal{R}_{j=2}^{(3)}=O(\epsilon^2)$.

\section{Correlators}\label{sect:correlators}
%
\begin{figure}[t]
\begin{picture}(200,70)(-10,0)
\put(10,0){\includegraphics[width=.40\textwidth]{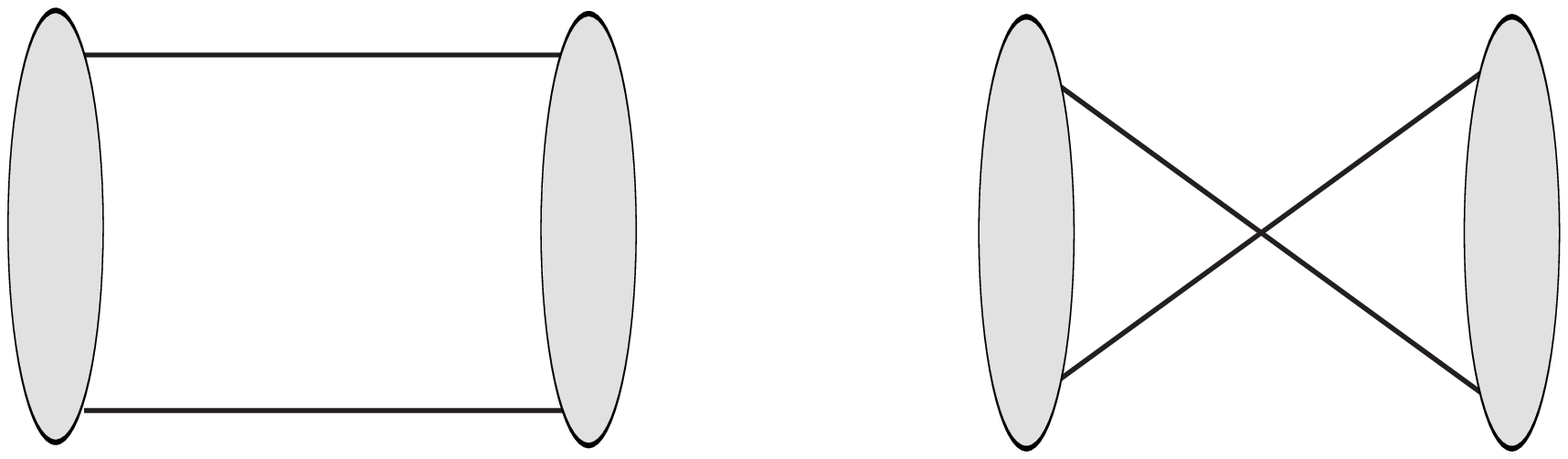}}
\put(2,5){$z_2$}
\put(2,50){$z_1$}
\put(93,5){$w_2$}
\put(93,50){$w_1$}
\put(122,5){$z_2$}
\put(122,50){$z_1$}
\put(207,5){$w_2$}
\put(207,50){$w_1$}
 \end{picture}
\caption{ The leading order diagrams for the correlator of two conformal operators, $\vev{\mathcal{O}^{(n)}_j(x)\mathcal{O}_j^{(\bar n)}(0)}$.
}
\label{fig:c220}
\end{figure}

\subsection{LO correlators}
In order to give a glimpse of the technique we start with calculation of the necessary correlators at leading order.
 The correlator
$\vev{\mathcal{O}^{(n)}_j(x)\mathcal{O}_j^{(\bar n)}(0)}$ is given by the sum of two diagrams, shown schematically in
Fig.~\ref{fig:c220}.
They  are given by the product of the propagators and give rise to  identical contributions to the correlation
function.
 Assuming that $x$ is chosen  in a transverse plane, $(x,n)=(x,\bar n)=0$, one represents the propagator as
\begin{align}\label{DK}
D(x+z_1n-\bar w_1 \bar n)=D(x){(1-z_1\bar w_1 r)^{-(\mu-1)}}\,,
\end{align}
where $r=2(n\bar n)/x^2$ and
\begin{align}
D(x)=\Gamma(\mu-1)\,/(4\pi^{\mu}{(x^2)^{\mu-1}})\,.
\end{align}
At leading order one replaces $\mu-1\mapsto2$  so that the second factor in~(\ref{DK}) is
nothing else as the reproducing kernel, $\mathcal{K}_{s=1}(z_1, w_1 r)$,
corresponding to the spin $s=1$, see Eq.~(\ref{r-kernel}).
Therefore starting from Eq.~(\ref{scalar-O}) one gets
for the correlator
\begin{align}\label{OO-tree}
\vev{\mathcal{O}^{(n)}_j(x)\mathcal{O}_j^{(\bar n)}(0)}&=2\xi D^2(x)(z_{12}^j| \prod_{k=1}^2\mathcal{K}_1(z_k, w_k r)|w_{12}^j)
\notag\\
&=2\xi D^2(x) r^j\, ||w_{12}^j||^2_{11}\,,
\end{align}
where  $\xi=n^2-1$ is the isotopic factor, the scalar products correspond to the conformal spin $s=1$  and
we take into account the property of the reproducing kernel~(\ref{fKf}).
The norm of $w_{12}^j$ is given by the following expression
\begin{align*}
||w_{12}^j||^2_{s_1s_2}=j!\prod_{k=1}^2\frac{\Gamma(2s_k)}{\Gamma(j+2s_k)}\frac{\Gamma(2j+2(s_1+s_2)-1)}{\Gamma(j+2(s_1+s_2)-1)}\,.
\end{align*}

\begin{figure}[t]
\begin{picture}(200,70)(-10,0)
\put(10,0){\includegraphics[width=.40\textwidth]{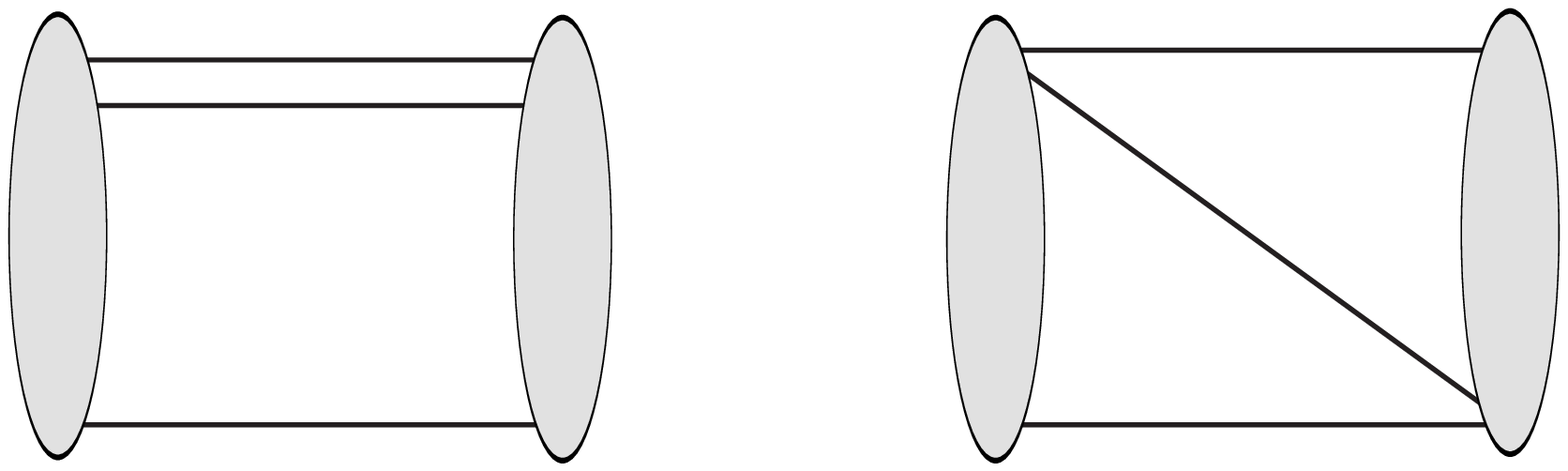}}
\put(2,5){$z_2$}
\put(2,50){$z_1$}
\put(91,5){$w_2$}
\put(91,50){$w_1$}
\put(120,5){$z_2$}
\put(120,50){$z_1$}
\put(207,5){$w_1$}
\put(207,50){$w_2$}
 \end{picture}
\caption{ The LO diagrams for the correlator of divergence of conformal operators, $\vev{\partial\mathcal{O}^{(n)}_j(x)\partial\mathcal{O}_j^{(\bar n)}(0)}$.
}
\label{fig:c330}
\end{figure}

The diagrams for the correlator $\vev{\partial\mathcal{O}^{(n)}_j(x)\partial\mathcal{O}_j^{(\bar n)}(0)}$ are shown
in Fig.~\ref{fig:c330}. On the leftmost diagram  the points $z_1n$ and $\bar w \bar
n$ are connected by two propagators 
\begin{align}\label{DK2}
D^2(x+z_1n-\bar w_1 \bar n)&=D^2(x){(1-z_1\bar w_1 r)^{-2(\mu-1)}}
\notag\\
&=D^2(x)\mathcal{K}_{s=2}(z_1, w_1r)\,.
\end{align}
Since $\partial \mathcal{O}_j^{(n)}\sim 
(z_{12}^j, L_{21} [\mathcal{O}(x,z_1,z_1,z_2)])_{11}$, see Eq.~(\ref{R23}),
the $z-$ scalar product has the form
\begin{align}\label{zsc}
(z_{12}^j, L_{21} \mathcal{K}_2(z_1, w_1 r)\mathcal{K}_1(z_2, w_2 r))_{11}\,.
\end{align}
The spins of the reproducing kernels and spins of the scalar product are in discord with each other. However, the operator
$L_{21}=\partial_2 z_{21}^2$  removes this mismatch. It intertwines the representations,
$$
L_{21} D_2^+\otimes D_1^+= D_1^+\otimes D_1^+ L_{21}
$$
and it can be easily shown, see e.g.
Ref.~\cite{Braun:2011dg},
that
\begin{align}
(z_{12}^j, L_{21} \Phi(z_1,z_2))_{11}=-a_j (z_{12}^{j-1}, \Phi(z_1,z_2))_{21}\,,
\end{align}
where
\begin{align}
a_j=(j+1){||z_{12}^{j}||_{11}^2}/{||z_{12}^{j-1}||_{21}^2}= \frac{j(j+2)(j+3)}6.
\end{align}
Thus the scalar product (\ref{zsc}) takes the form
\begin{align}\label{zsc1}
-a_j(z_{12}^{j-1},  \mathcal{K}_2(z_1, w_1 r)\mathcal{K}_1(z_2, w_2 r))_{21}=-a_j (r \bar w_{12})^{j-1}.
\end{align}
Restoring all
color and symmetry factors one gets for the first diagram
\begin{multline}\label{OO3-tree}
-\frac12 \xi \lambda_s g^2 D^3(x) a_j/(j+1)^2r^{j-1} (L_{21}\bar w_{12}^{j-1}, w_{12}^j)_{11}
=\\
=\frac12\xi \lambda_s g^2 D^3(x) a_j/(j+1) r^{j-1} ||w_{12}^j||_{11}^2\,.
\end{multline}

The calculation of the second diagram goes along the same line. One can combine propagators attached to the point
$z_1$ using the Feynman's trick to get
\begin{multline}
(z_{12}^j, L_{21} \mathcal{K}_2(z_1, w_1 r)\mathcal{K}_2(z_1, w_2 r)\mathcal{K}_1(z_2, w_1 r))_{11}=
\\
=\frac{6(-1)^ja_j (r \bar w_{12})^{j-1}} {(j+1)(j+2)}\,.
\end{multline}
Finally, taking into account that second diagram enters with symmetry factor $2$ one gets,  in full agreement
with~(\ref{OO/OO}),
\begin{align}\label{TLO}
\mathcal{T}_j(u_*)&=u_*\varkappa_j \gamma_j^{(1)}+O(u_*^2)\,,
\end{align}
where  the one--loop anomalous dimension is, see Eq.~(\ref{H-one-loop}),
\begin{align}\label{gamma_1}
\gamma_j^{(1)}=\lambda_s\frac16\frac{(j-2)(j+5)}{(j+1)(j+2)}\,.
\end{align}
Thus the diagrams are easily calculated provided that the spins of the reproducing kernels match that of the scalar
product. We will show that this scheme can be extended to loop diagrams as well.

\subsection{NLO correlators}
%
\begin{figure*}[t]
\begin{picture}(600,100)(-20,-20)
\put(10,10){\includegraphics[width=0.85\textwidth]{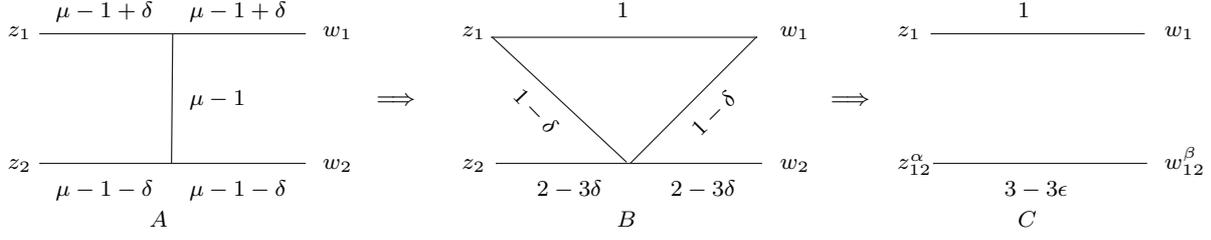}}
\put(2,10){$z_2$}
\put(2,60){$z_1$}

\put(120,10){$w_2$}
\put(120,60){$w_1$}

\put(140,34){$\Longrightarrow$}
\put(172,10){$z_2$}
\put(172,60){$z_1$}

\put(291,10){$w_2$}
\put(291,60){$w_1$}

\put(310,34){$\Longrightarrow$}

\put(335,10){$z_{12}^\alpha$}
\put(335,60){$z_1$}

\put(435,10){$w_{12}^\beta$}
\put(435,60){$w_1$}

\put(20,0){$\mu-1-\delta$}
\put(70,0){$\mu-1-\delta$}
\put(55,-12){$A$}

\put(20,67){$\mu-1+\delta$}
\put(70,67){$\mu-1+\delta$}

\put(70,34){$\mu-1$}

\put(200,0){$2-3\delta$}
\put(250,0){$2-3\delta$}
\put(230,-12){$B$}

\put(230,67){$1$}
\put(190,37){\rotatebox{-46}{$1-\delta$}}
\put(257,22){\rotatebox{+45}{$1-\delta$}}

\put(380,67){$1$}
\put(375,0){$3-3\epsilon$}
\put(380,-12){$C$}
 \end{picture}
\caption{NLO correction  to the correlator of conformal operators $\vev{\mathcal{O}^{(n)}_j(x)\mathcal{O}_j^{(\bar n)}(0)}$.
The parameter $\delta=\epsilon/2$.}
\label{fig:c221}
\end{figure*}

First, it is easy to see that the corrections due to modification of the scalar product cancel out in the ratio of
correlators.
Indeed, these corrections only influence  the norm
$$
||w_{12}^j||_{11}^2\mapsto ||w_{12}^j||_{\varpi}^2=(w_{12}^j, (1+\varpi^{(1)})w_{12}^j)_{11}.
$$
entering the tree level expressions,  Eqs.~(\ref{OO-tree}), (\ref{OO3-tree}), etc., which
cancels out in the ratio of correlators, $\mathcal{T}_j$, irrespectively of the
explicit form of $\varpi$. This cancellation is expected. Indeed,  the problem can be reformulated as a standard  quantum mechanical
problem for a certain Hamiltonian. A modification of the scalar product produces  corrections to the eigenstates (conformal
operator). However,  the energy shift at the leading order, $\delta
E^{(1)}_\psi=\vev{\psi^{(0)}|V|\psi^{(0)}}$, is not sensitive to such corrections.

The calculation of $\vev{\mathcal{R}_j^{(2)}\mathcal{R}_j^{(3,0)}}$ is a bit more  involved
but  straightforward. The corresponding contribution to $T_j^{(2)}$ reads
\begin{align}\label{mixed}
 T_j^{(2\to3)}=-\frac12(\gamma_j^{(1)})^2\,
\frac{(j-1)(j^2+4j+9)}{(j+1)(j+2)(j+3)(j+5)}.
\end{align}
It is convenient to split  one loop corrections to the correlators
 $\vev{\mathcal{O}_j \mathcal{O}_j}$ and  $\vev{\mathcal{R}_j^{(3,0)}\mathcal{R}_j^{(3,0)}}$
into two groups
\begin{itemize}
\item Self-energy insertions to the propagators.
\item All other loop diagrams
\end{itemize}

Taking into account the self-energy correction to the propagators is equivalent to the calculation of the tree--level
diagrams with the exact (critical) propagator
\begin{align}
D_c(x)=A(u_*)/(x^2)^{\Delta_\varphi}\,,
\end{align}
where $\Delta_\varphi=\mu-1+\gamma_\varphi$ is the critical dimension of the basic field and
the residue $A(u_*)$ is
\begin{align}
A(u_*)=\frac{\Gamma(\mu-1)}{4 \pi^\mu\overline{M}^{2\gamma_\varphi}}\left(1-u_*\lambda_s\frac5{36}+O(u_*^2)\right),
\end{align}
where $\overline{M}^2=\pi M^2e^{\gamma_E}$. 
For the first diagram in Fig.~\ref{fig:c220} one gets
\begin{align}\label{Dcsc}
D_c^2(x)\,\, r^j\,\, (z_{12}^j|\mathcal{K}_{s}(z_1,w_1)\mathcal{K}_{s}(z_2,w_2)|w_{12}^j)_{(11),(11)}\,,
\end{align}
where $s=\Delta_\varphi/2=1-(\epsilon-\gamma_\varphi)/2$, the subscripts indicate the conformal spins of the
$z$  and $w$ scalar products, respectively. The  reproducing kernels in~(\ref{Dcsc})  correspond to  spin $s$ which
does not match the spins of the scalar product. However,
it is easy to see  that, due to symmetry, the scalar product with  modified spins
\begin{align}\label{Dss}
(z_{12}^j|\mathcal{K}_{s}(z_1,w_1)\mathcal{K}_{s}(z_2,w_2)|w_{12}^j)_{(s_+^\delta,s_+^\delta),(s_-^\delta,s_-^\delta)}\,,
\end{align}
where $s^\delta_\pm=1\pm \delta$
is equal to that in~(\ref{Dcsc}) up to terms of order $\delta^2$. Therefore, choosing $s_-^\delta=s$,
($\delta=(\epsilon-\gamma_\varphi)/2$),  and  evaluating the $w-$product in~(\ref{Dss}) one
gets for (\ref{Dcsc})
\begin{align}
D_c^2(x)\,\, r^j\,||z_{12}^j||_{s_+^\delta,s_+^\delta}^2+O(u_*^2)\,.
\end{align}
The calculation of the diagrams in Fig.~\ref{fig:c330} goes  along the same lines.  Finally,  the contribution to
the ratio of correlators $\mathcal{T}_j$ from
the leading order diagrams and
self-energy diagrams   can be written in the form (up to $O(\epsilon^2)$ terms)
%
\begin{align}\label{SE}
\mathcal{T}_j^{SE}&=\frac{2u_*\lambda_s}{\varkappa_j}\left(1+\frac{u_*}{2}\left(\lambda_d-\frac7{9}\lambda_s\right)\right)
\frac{||z_{12}^{j-1}||^2_{2s_+^\delta,s_+^\delta}}{||z_{12}^{j}||^2_{s_+^\delta,s_+^\delta}}
\notag\\
&\quad\times\left(1-\frac{2\Gamma(4s_-^\delta)(\Gamma(j-1+2s_-^\delta)}{\Gamma(2s_-^\delta)\Gamma(j-1+4s_-^\delta)}\right).
\end{align}
Expanding~(\ref{SE}) we find for the corresponding contribution to the coefficient $T^{(2)}_j$
\begin{align}\label{TSE}
{T}^{SE}_j&=\gamma_j^{(1)}\Biggl\{\frac{1}{2}\left(\lambda_d-\frac7{9}\lambda_s\right)
\notag\\
&+2\delta\Biggl[S_{2j+2}
-S_{j+3}-S_j +\frac23\frac{4j^2+14j+9}{(j+2)(j+3)}\Biggr]\Biggr\}
\notag\\[2mm]
&
-\frac{4 \lambda_s\,\delta}{(j+1)(j+2)}\left[S_{j+2}-\frac{(2j+1)(4j+7)}{3(j+1)(j+2)}\right],
\end{align}
where $S_j=\sum_{k=1}^j 1/k$ and
\begin{align}
\delta
=\frac{1}{4}\left(-\lambda_d+\frac13{\lambda_s}\right)\,.
\end{align}
We recall that 
$\gamma_j^{(1)}$
is the one loop anomalous dimension, Eq.~(\ref{gamma_1}).

\subsection{Loop diagrams}

All loop diagrams can be calculated quite easily with the help of several simple tricks. Several examples are given
below.
Let us start with a correction to the correlator of two conformal operators.
The corresponding contribution has the form
\begin{align}\label{Cepsilon}
C(\epsilon)=2\,(z_{12}^j| H(x;z,w)|w_{12}^j)_{(11),(11)}\,,
\end{align}
where the kernel $H(x;z;w)$ is given by the left diagram shown in Fig.~\ref{fig:c221} with the parameter $\delta\to 0$.
We need to find $C(\epsilon)$ up to terms $O(\epsilon^0)$ , $C(\epsilon)=\frac1\epsilon(c_0+\epsilon
c_1+\ldots)$. To this end  we proceed as follows. We modify the indices as shown in Fig.~\ref{fig:c221}. This
modification does
not change the pole structure (see discussion in Ref.~\cite{Ciuchini:1999wy}) and, due to the symmetry
$C(\epsilon,\delta)=C(\epsilon,-\delta)$, one concludes that
\begin{align}\label{Cexp}
C(\epsilon,\delta)=\frac1\epsilon(c_0+\epsilon c_1+c_2\delta^2+\ldots)\,.
\end{align}
The choice  $\delta=\epsilon/2$ results in
the uniqueness of the upper integration vertex  and at the same time  does not affect
first two terms in~\eqref{Cexp}. Using the star--triangle relation for the upper vertex one gets
the diagram $B$ in Fig.~\ref{fig:c221}.  Using Feynman formula for the left (right) propagators attached to the
integration vertex one can perform the last integral that results in the  diagram $C$ in Fig.~\ref{fig:c221}.

\begin{figure}[h]
\begin{picture}(200,100)(10,-5)
\put(50,0){\includegraphics[width=.55\columnwidth]{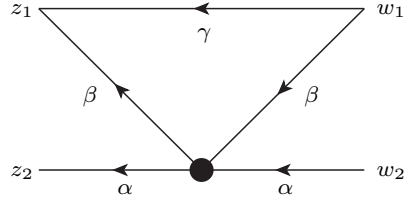}}
\put(45,5){${z_2}$}
\put(45,66){$z_1$}
\put(182,5){$w_2$}
\put(182,66){$w_1$}

\put(85,-2){$\displaystyle\alpha$}
\put(145,-2){$\displaystyle\alpha$}

\put(72,33){$\displaystyle\beta$}
\put(155,33){$\displaystyle\beta$}

\put(115,57){$\displaystyle\gamma$}
 \end{picture}
\caption{ The ``sl(2)'' diagram: an arrow line from $w$ to $z$ with index $\alpha$ stands for the propagator
$(1-z\bar w)^{-\alpha}$.
The indices have the following values: $\alpha=2-3\epsilon/2$, $\beta=1-\epsilon/2$ and $\gamma=1$.
 The black circle denote an integration vertex with the $sl(2)$ invariant measure $\mu_{s+\delta}$,
 $s+\delta=3/2-\epsilon/4$.}
\label{fig:sl2d}
\end{figure}

This diagram, up to a $x-$dependent factor, has the form
\begin{align}\label{KK2}
  \mathcal{K}_{\frac12}(z_1,w_1)\int_0^1d\alpha d\beta (\bar\alpha\bar\beta)^{-\delta}(\alpha\beta)^{1-3\delta}
\mathcal{K}_{\frac32-3\delta}(z_{12}^\alpha,w_{12}^\beta)\,.
\end{align}
On the next step we want to get rid of the parametric integrals. To this end we use the properties of the reproducing
kernel~(\ref{fKf}) and represent
\begin{multline}\label{KKN}
\mathcal{K}_{s-\delta}(z_{12}^\alpha,w_{12}^\beta)=\\
\int d^2\xi'\, \mu_{s}(\xi')\mathcal{K}_{s}(z_{12}^\alpha,\xi')
 \int d^2\xi\, \mu_{s}(\xi)\mathcal{K}_{s-\delta}(\xi',\xi)\,
\mathcal{K}_{s}(\xi,w_{12}^\beta)
\\
=\int d^2\xi\, \mu_{s+\delta}(\xi)\mathcal{K}_{s}(z_{12}^\alpha,\xi)
\mathcal{K}_{s}(\xi,w_{12}^\beta)
+O(\delta^2)\,,
\end{multline}
where $s=3/2-2\delta$ and we used the  relation~(\ref{Kmu}). Using this expression one can carry out the
integrals over $\alpha,\beta$ in~(\ref{KK2}). The resulting expression for $H(x;z,w)$ (up to a prefactor)
takes the form of  ``$sl(2)$'' diagram shown in Fig.~\ref{fig:sl2d}.

\begin{figure*}[t]
\begin{picture}(600,90)(-20,-10)
\put(10,10){\includegraphics[width=0.85\textwidth]{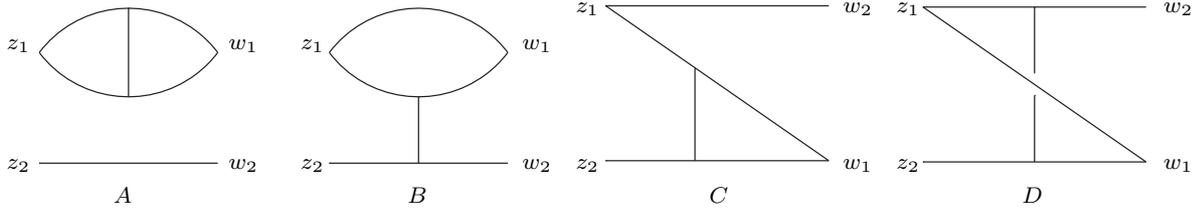}}
\put(2,10){$z_2$}
\put(2,55){$z_1$}

\put(85,10){$w_2$}
\put(85,55){$w_1$}

\put(112,10){$z_2$}
\put(112,55){$z_1$}

\put(195,10){$w_2$}
\put(195,55){$w_1$}

\put(215,10){$z_2$}
\put(215,69){$z_1$}

\put(315,10){$w_1$}
\put(315,69){$w_2$}

\put(335,10){$z_2$}
\put(335,69){$z_1$}

\put(435,10){$w_1$}
\put(435,69){$w_2$}

\put(42,-3){$A$}
\put(152,-3){$B$}

\put(265,-3){$C$}
\put(382,-3){$D$}

 \end{picture}
\caption{One loop correction to the correlator of the divergence of conformal operators $\vev{\partial\mathcal{O}^{(n)}_j(x)\partial\mathcal{O}_j^{(\bar
n)}(0)}$.}
\label{fig:c33A1}
\end{figure*}
Now we need to calculate the scalar product (\ref{Cepsilon}) with the kernel $H(x;z,w)$ given by the diagram on
Fig.~\ref{fig:sl2d}. The scalar product is a function of indices of left (right) propagators and conformal spins of $z$
($w $) scalar products,
$
S(a,b, s_1,s_2|a', b',s'_1,s'_2).
$
We need this function up to $O(\epsilon)$ terms   for $a=a'=\alpha$, $b=b'=\beta$ and $s_i=s'_i=1$. As was explained
in the previous section the integral  with measure $\mu_s$ can be  evaluated  easily provided that  the sum of the indices of
propagators coming from  this vertex is equal to $2s$. Taking this into account one finds that the scalar product with shifted indices
$$
S(\alpha+\delta,\beta+\delta, 1+2\delta,1+4\delta |\alpha-\delta,\beta-\delta, 1-2\delta,1-4\delta )
$$
can be straightforwardly calculated to
\begin{align}
S(\epsilon)=\frac{j!\Gamma(3-\epsilon)}{\Gamma(j+3-\epsilon)}\,  ||z_{12}^{j}||^2_{1+\frac12\epsilon,1+\epsilon}
\end{align}
and in the same time it differs from the scalar product in question by terms of order $O(\epsilon^2)$ only.

Restoring all factors one gets
for $C(\epsilon)$
\begin{align}
C(\epsilon)&=u_* \, \lambda_s (n^2-1) D^2(x) (x^2\overline{M}^2)^\epsilon\, {r^{j}}
\frac{2+\epsilon}\epsilon\, S(\epsilon)\,.
\end{align}
Since the ratio~(\ref{OO/OO}) does not depend on $x$, it is convenient to put $x^2\overline{M}^2=1$.
Finally, subtracting the counterterms
\begin{align*}
\Delta C(\epsilon)=-\frac1\epsilon4u_* \, \lambda_s (n^2-1) D^2(x)\, {r^{j}}
\frac{||z_{12}^{j}||^2_{1+\frac12\epsilon,1+\frac12\epsilon}}{(j+1)(j+2)}\,,
\end{align*}
one obtains
\begin{align}\label{OO-loop}
\frac{C(\epsilon)+\Delta C(\epsilon)}{\VEV{\mathcal{O}_j(x)\mathcal{O}_j(0)}_{0}}={u_*\lambda_s}\frac{2(S_{2j+2}-S_{j+1})}{(j+1)(j+2)}+\cdots
\end{align}
where ellipses stand for higher order terms.
The  corresponding contribution from  (\ref{OO-loop})  to the coefficient  $T_j^{(2)}$ in the ratio of the correlators,
see Eq.~\eqref{OO/OO},
reads
\begin{align}
T_j^{(o)} = - \gamma_j^{(1)} {\lambda_s}\frac{2(S_{2j+2}-S_{j+1})}{(j+1)(j+2)}\,.
\end{align}
\vskip 0.4cm

\begin{figure}[t]
\begin{picture}(200,80)(0,-10)
\put(0,00){\includegraphics[width=0.99\columnwidth]{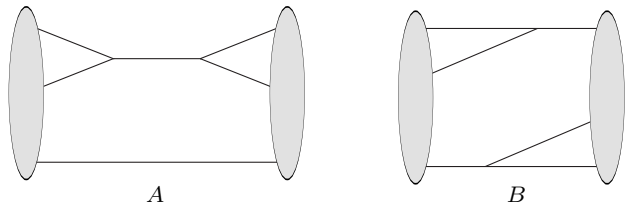}}
\put(55,-5){$A$}
\put(190,-5){$B$}
 \end{picture}
\caption{ NLO diagrams for the correlator of the divergence of conformal operators
$\vev{\partial\mathcal{O}^{(n)}_j(x)\partial\mathcal{O}_j^{(\bar n)}(0)}$.}
\label{fig:c33B1}
\end{figure}

The NLO diagrams contributing to the correlator   $\VEV{\partial{\mathcal{O}_j}(x)\partial{\mathcal{O}_j}(0)}$
are shown in Fig.~\ref{fig:c33A1} (type A) and Fig.~\ref{fig:c33B1} (type B). These diagrams have different color
factors: $\lambda_s\lambda_d$ for the A - diagrams, and $\lambda_s^2$ for the B - diagrams.

All diagrams of the A - type  have  the diagram we have just now discussed as a subgraph.
The only difference is that in order to kill  terms linear in $\delta$
one has to consider an average of the diagrams,
$(D(\delta)+D(-\delta))/2$. For $\delta=\epsilon/2$ each of the diagrams, $D(\pm\delta)$, can be simplified with the help
of the star--triangle relation and rewritten in the form of ``$sl(2)$'' diagrams. These diagrams in turn can be
calculated up to $O(\epsilon^2)$ terms in a manner  described above. So we skip all details and present the result
for  each diagram in \ref{app:D}.

All  diagrams of B-type shown schematically in Fig.~\ref{fig:c33B1}   contain two  $2\to 1$ subgraphs. The diagrams depicted on
the right panel are finite while those on the left panel are divergent. The calculation  of these diagrams does not
present any problem so that we give only final results in  \ref{app:D}.

Finally, it follows from Eq.~(\ref{LZ=ZL}) that the sum of counterterm diagrams  to the diagrams in Figs.~\ref{fig:c33A1}
and \ref{fig:c33B1} can be written in the form
\begin{align}
2(Z_3 Z_1^{-1}Z_{j}-1)\,\vev{\partial\mathcal{O}_j^{(n)}(x)\partial\mathcal{O}_j^{(\bar n)}(0)}^{(\epsilon)}_0\,.
\end{align}
Here $Z_j$ is the one--loop renormalization constant for the operator $\mathcal{O}_j$,
\begin{align}
Z_j=1-\frac{u}{\epsilon}\frac{\lambda_s}{(j+1)(j+2)}
\end{align}
and
\begin{align}
Z_1=1-\frac{u\lambda_s}{12\epsilon}+O(u^2),
&&
Z_3=1-\frac{u\lambda_d}{4\epsilon}+O(u^2)\,.
\end{align}
We have put the superscript $(\epsilon)$ to  the correlator in order to stress that even the tree--level correlator depends on
$\epsilon$ through a space-time dimension $d=6-2\epsilon$.

\section{Results}\label{sect:results}
The coefficient $T_j^{(2)}$ in the ratio of the correlators is given by  the sum  of terms in
Eqs.~(\ref{mixed}), (\ref{TSE}), (\ref{OO-loop}), (\ref{T5}) and (\ref{T6}).
In the case of operators with other isotopic symmetry these expressions have to  be modified.
One can separate seven different isotopic projections
\begin{align}
\mathcal{O}_f^{ab}(x;z_1,z_2)=(P_{f})^{ab}_{a'b'}\varphi^{a'}(x+z_1n)\varphi^{b'}(x+z_2n),
\end{align}
where $f=1,\ldots,7$. The  projectors $P_f$ can be found in Ref.~\cite{Braun:2013tva}.
The one--loop anomalous dimension for  the operator $\mathcal{O}_j^{(f)}$ is given by
\begin{align}\label{gammaf1}
\gamma_j^{f(1)}=\frac16\left(\lambda_1-\frac{12\lambda_f}{(j+1)(j+2)}\right)\,,
\end{align}
where $\lambda_f$ are the eigenvalues of the  operator $\mathbb{R}^{ab}_{a'b'}=d^{aa'c}d^{bb'c}$ on the invariant
subspaces, $\mathbb{R}P_f=\lambda_f P_f$. The explicit expressions for $\lambda_f$ can be found
in~\cite{Braun:2013tva}. We note also that  $\lambda_s=\lambda_1$ and $\lambda_d=2\lambda_3$.

The modifications of the expressions~(\ref{mixed}), (\ref{TSE}), (\ref{OO-loop}), (\ref{T5}) and (\ref{T6}) for the
case of  arbitrary projections, $P_f$, are the following:
\begin{itemize}
\item Replace $\gamma_j^{(1)}\to \gamma_j^{(1,f)}$ in all expressions.

\item  Replace $\lambda_s\to \lambda_f$ in the expressions for $T^{(o)}_j$, $T^{(5,B)}_j$, $T^{(5,C)}_j$,  $T^{(6,A)}_j$
and  in the last line of $T^{SE}_j$, Eq.~(\ref{TSE}).

\item Replace $\lambda_s\lambda_d\to 2\nu_f$ in the expression for $T^{(5,D)}_j$,
\end{itemize}
where $\nu_f$ are the eigenvalues of the invariant operator $\mathbb{T}^{ab}_{a'b'}=(\mathbb{R}^2)^{a, b'}_{a'b}$,
see Ref.~\cite{Braun:2013tva}.
%

Representing the ratio of the correlators in the form
\begin{align}\label{Tf12}
\mathcal{T}^{f}_j(u_*) & =\varkappa_j\Big(u_* \,T_j^{f(1)} ~ +~ u_*^2\, T_j^{f(2)} +  \ldots\Big)
\end{align}
one obtains for the coefficient $T_j^{f(2)}$:
\begin{align}
T_j^{f(2)} & = \sum_{ab} \lambda_a\,\lambda_b \,T^{(2)}_{j,ab}+\nu_f\, T^{(2)}_{j,f}
\end{align}
where
\begin{align}
T^{(2)}_{j,ss}&=-\frac{1}{72}\left(\frac{11}3+\frac{2j^2+7j-1}{(j+1)(j+2)(j+3)}\right)\,,
\notag\\[1mm]
T^{(2)}_{j,sd}&=\frac{1}{12}\left(\frac23+\frac{2j^2+5j+1}{(j+1)(j+2)(j+3)}\right)\,,
\notag\\[1mm]
T^{(2)}_{j,ff}&=2\frac{j^3+8j^2+10j+1}{(j+1)^3(j+2)^3(j+3)}\,,
\notag\\[1mm]
T^{(2)}_{j,fd}&=\frac{1}{(j+1)(j+2)}\left( S_{j+1}-3+\frac4{(j+1)(j+3)}\right)\,,
\notag\\[1mm]
T^{(2)}_{j,fs}&=-\frac{1}{3(j+1)(j+2)}\biggl(S_{j+3}-4
\notag\\
&\quad
\hskip 22mm +\frac{j^2+2j+5}{(j+1)(j+2)(j+3)}\biggr)\
\end{align}
and
\begin{align}
T^{(2)}_{j,f}&=-\frac{2}{(j+1)^2(j+2)^2}\,.
\end{align}
Finally, comparing \eqref{Tf12} with the r.h.s. of Eq.~\eqref{OO/OO} we get for the anomalous dimension
\begin{align}
\gamma_j^f(u_*)&=u_*\gamma_j^{f,(1)}+u_*^2\gamma_j^{f,(2)}+\ldots,
\end{align}
where $\gamma_j^{f(1)}$ is given by Eq.~(\ref{gammaf1}) and  $\gamma_j^{f(2)}$ has the form
\begin{align}
\gamma_j^{f(2)}&=2\gamma_{\varphi}^{(2)}+\frac{2\nu_f}{(j+1)^2(j+2)^2}
+2\lambda_f^2\frac{j^2+j-1}{(j+1)^3(j+2)^3}
\notag\\[2mm]
&\quad
-\frac{\lambda_f\lambda_s}{3(j+1)(j+2)}\left[S_{j+2}-4+\frac{2j+3}{2(j+1)(j+2)}\right]
\notag\\[2mm]
&\quad
+\frac{\lambda_f\lambda_d}{(j+1)(j+2)}\left[ S_{j+2}-3+\frac1{j+1}\right]\,.
\end{align}
This expression completely agrees with  the anomalous dimensions reconstructed from the two--loop  evolution
kernels~\cite{Braun:2013tva}.
We have also checked that the large $j$ expansion of the anomalous dimensions $\gamma_j^f(u_*)$ in terms  of
the quadratic Casimir
$$
J^2=\left(j+2-\epsilon+\frac12\gamma_j^f(u_*)\right)\left(j+1-\epsilon+\frac12\gamma_j^f(u_*)\right)
$$
contains only even powers of $1/J$ \cite{Basso:2006nk,Alday:2015eya}.

\section{Summary}\label{sect:summary}
We have calculated two--loop anomalous dimensions of the leading--twist operators in the $\varphi^3$ model using the
approach proposed in Refs.~\cite{Anselmi:1998ms,Belitsky:2007jp}.  Formally this me\-thod allows one to gain one order
in the perturbation theory. However, this advantage is illusory since one has to calculate diagrams including finite
parts instead of the pole terms in the standard approach. There is  also no essential gain in a complexity of
calculations.

Nevertheless, both the contributing diagrams   and the methods of  calculation
are quite different in the two approaches. Therefore the calculation of anomalous dimensions performed in this approach
could provide an additional check of the results obtained within the standard approach. Of course, going to the next order
is  only possible  with the advanced methods of computer algebra.

\begin{acknowledgements}
We are grateful to Vladimir Braun for useful discussions.
This work has been supported by Deutsche Forschungsgemeinschaft through grant MO 1801/1-1.
\end{acknowledgements}

\appendix
\setcounter{equation}{0}
\section{Discrete series representations of the  $su(1,1)$ group}
\label{app:A}
At the leading order the light-ray
operator $\mathcal{O}(x;z_1,z_2)$ transforms according to the tensor product of discrete series representations of the
group $su(1,1)$. In this Appendix we  recall   their basis properties.
The discrete series representation of
the $su(1,1)$ group, $D^+_s$ in the standard notations~\cite{GelGraVil66}, is defined on the space of functions analytic inside
the unit circle, $|z|<1$,
\begin{align}
D_{s}^+(g^{-1}) f(z)=(\bar b z+\bar a)^{-2s} f(z')\,,
\end{align}
where $z'=(az+b)/(\bar b z+\bar a)$, and $g=\begin{pmatrix}
a & b
\\
\bar b &\bar a
\end{pmatrix} \in su(1,1)$
and $s$ is an integer or half-integer.
The generators of the group have the form
\begin{align}
S_-=-\partial_z,&& S_0=z\partial_z+s\,, && S_+=z^2\partial_z+2s z\,.
\end{align}
The invariant scalar product~\cite{GelGraVil66,Vilenkin}
$$(D_{s}^+(g) f,D_{s}^+(g) \psi)_s=(f,\psi)_s$$
is uniquely defined (up to unitary equivalence)
\begin{align}\label{scs}
(f,\psi)_s=\int d^2z\,\mu_s(z)\, \overline{f(z)}\,\psi(z)\,,
\end{align}
where
$$
\mu_s(z)=\dfrac{2s-1}\pi(1-|z|^2)^{2s-2}\theta(1-|z|).
$$
The space of analytic functions on the unit disk with the scalar product~(\ref{scs}) is called a holomorphic Hilbert
space, $D_s^+$, see for a review Ref.\,\,\cite{Hall}. The powers of $z$ form an orthogonal basis, $\{e_k(z)=z^k,\  k=0,1,\ldots\}$, in this space,
\begin{align}
(e_{m},e_k)_s=\delta_{km} ||e_k||^2_s=\delta_{mk}\frac{\Gamma(2s) k!}{\Gamma(k+2s)}\,.
\end{align}
The unit operator (the reproducing kernel) has the form
\begin{align}\label{r-kernel}
\mathcal{K}_s(z,w)=\sum_{k=0}^\infty\! e_k(z)\overline{e_k(w)}/||e^k||^2_s={(1-z\bar w)^{-2s}}
\end{align}
and for an arbitrary function  $f\in D_s^+$ the following identity holds
\begin{align}\label{fKf}
f(z)=\int d^2w\,\mu_s(w)\,\mathcal{K}_s(z,w) \,f(w)\,.
\end{align}
We note here that all  formulae (\ref{scs}) -- (\ref{r-kernel}) have a perfect sense for any $s\geq 1/2$.

Finally, we give a  relation that turned out to be  very useful in the calculations
\begin{multline}\label{Kmu}
\int d^2w\,\mu_{s+\epsilon}(w)\,\mathcal{K}_s(z,w) \,f(w)=
\\
=\int  d^2w\,\mu_{s}(w)\,\mathcal{K}_{s-\epsilon}(z,w) \,f(w)+O(\epsilon^2)\,.
\end{multline}
It follows immediately from~(\ref{fKf}) if one replaces $s\to s+\epsilon$ and expands it in $\epsilon$.

\setcounter{equation}{0}
\section{One loop scalar product}
\label{app:B}
The one loop correction $\varpi^{(1)}$  to the scalar product~(\ref{sc}) is determined by
Eqs.~(\ref{SDelta}). To find the solution we note that the one loop kernel $\mathcal{H}^{(1)}$ can be represented in the
factorized form~\footnote{
Let us remark that the evoluion kernel $\mathcal{H}^{(1)}$ can be identified with the
$sl(2)-$invariant $\mathcal{R}-$operator for
a special value of  spectral parameter, $\mathcal{H}^{(1)}=\mathcal{R}_{s_1=1,s_2=1}(u=-i)$.
The factorization of $\mathcal{H}^{(1)}$ is a consequence  of a factorization property of
the $\mathcal{R}-$operator~\cite{Derkachov:2005hw}, see also Ref.~\cite{Derkachov:2005xn}.
}
\begin{align}\label{Hfac-form}
\mathcal{H}^{(1)}=\overline{\mathcal{F}}_{12} \mathcal{F}_{12}=\overline{\mathcal{F}}_{21} \mathcal{F}_{21}\,,
\end{align}
where
\begin{align}
\mathcal{F}_{12}=\left( \partial_2 z_{21}\right)^{-1}\,, &&\overline{\mathcal{F}}_{12}=\left(z_{12}^{-1}\partial_1 z_{12}^2\right)^{-1}\,,
\notag\\
\mathcal{F}_{21}=\left( \partial_1 z_{12}\right)^{-1}\,, &&\overline{\mathcal{F}}_{21}=\left(z_{21}^{-1}\partial_2 z_{21}^2\right)^{-1}\,.
\end{align}
In order to obtain~(\ref{Hfac-form}) it is sufficient to notice that $\mathcal{H}^{(1)}$ is nothing else as the
inverse Casimir operator,
$$
(\mathcal{H}^{(1)})^{-1}= -\partial_1\partial_2 z_{12}^2
=\mathcal{F}_{12}^{-1}\,\overline{\mathcal{F}}_{12}^{-1}
=\mathcal{F}_{21}^{-1}\,\overline{\mathcal{F}}_{21}^{-1}.
$$
While the operator $\mathcal{H}^{(1)}: D_+^{(1)}\otimes D_+^{(1)}\mapsto D_+^{(1)}\otimes D_+^{(1)}$, the
$\mathcal{F}$ operators intertwine the representations with different spins. Namely,
\begin{align}
\mathcal{F}_{12}\,D_+^{(1)}\otimes D_+^{(1)}&=D_+^{(\frac32)}\otimes D_+^{(\frac12)}\,\mathcal{F}_{12}\,,
\notag\\
\overline{\mathcal{F}}_{12}\,D_+^{(\frac32)}\otimes D_+^{(\frac12)}&=D_+^{(1)}\otimes D_+^{(1)}\,\overline{\mathcal{F}}_{12}\,,
\end{align}
and similar for $\mathcal{F}_{21}$. It results in the  intertwining relations for two-particle generators,
$S^{(s_1,s_2)}_\alpha=S_{\alpha}^{(s_1)}+S_{\alpha}^{(s_2)}$,
\begin{align}\label{FFS}
\mathcal{F}_{12}S^{(1,1)}_\alpha=S^{(\frac32,\frac12)}_\alpha\mathcal{F}_{12}\,, &&
\mathcal{F}_{21}S^{(1,1)}_\alpha=S^{(\frac12,\frac32)}_\alpha\mathcal{F}_{21}
\end{align}
and  so on.

Next, we introduce one particle operator
\begin{align}
W^{(s)}f(z)=\int_0^1d\alpha \frac{\bar\alpha^{2s-1}}\alpha
\Big(f(z)-f(\bar\alpha z)\Big)\,,
\end{align}
such that $W^{(s)} z^{n}=(\psi(n+2s)-\psi(2s)) z^{n}$. This operator commutes with the generator $S_0^{(s)}$ while
\begin{align}\label{SFz}
[S_+^{(s)}, W^{(s)}]=-z\,.
\end{align}
Now let us check that
\begin{align}
\varpi^{(1)}&=(\epsilon-u_*\gamma_\varphi^{(1)})\Big(W_1^{(1)}+W_2^{(1)}\Big)\notag\\
&
\quad+
\lambda_s u_*\Big(\overline{\mathcal{F}}_{12}W_2^{(\frac12)} \mathcal{F}_{12}+\overline{\mathcal{F}}_{21}W_1^{(\frac12)} \mathcal{F}_{21}\Big)
\end{align}
gives solution to  Eqs.~(\ref{SDelta}). The first equation is obviously satisfied, $[S_0^{(0)},\varpi^{(1)}]=0$.
Next, making use of  Eqs.~(\ref{FFS}), (\ref{SFz}) one gets for the commutator
\begin{align}
[S_+^{(0)},\varpi^{(1)}]&=-(\epsilon-u_*\gamma_\varphi^{(1)})(z_1+z_2)
\notag\\
&\quad
-\lambda_s u_*\Big(\overline{\mathcal{F}}_{12} z_2 \mathcal{F}_{12}+\overline{\mathcal{F}}_{21}z_1 \mathcal{F}_{21}\Big)\,.
\end{align}
Finally, one casts the  r.h.s. into the necessary form  taking into account that
$[\overline{\mathcal{F}}_{12},z_2]=[\overline{\mathcal{F}}_{21},z_1]=0$.

The  scalar product can also be written in the form
\begin{align}\label{scalar-exact}
(f,\psi)_\varpi&=(f,\psi)_{s_*,s_*}+\lambda_s u_* \biggl[(\mathcal{F}_{12}f,W_2^{(\frac12)}\mathcal{F}_{12}\psi)_{\frac32,\frac12}
\notag\\
&\quad
+(\mathcal{F}_{21}f,W_1^{(\frac12)}\mathcal{F}_{21}\psi)_{\frac12,\frac32}\biggr]+O(\epsilon^2)\,,
\end{align}
where $s_*=2-\epsilon+\gamma_\varphi^*$ is the conformal spin of the basic field  at the critical point
and $(f,\psi)_{s_1,s_2}$ stays for the two particle scalar product. We have to mention here that the solution of
Eqs.~(\ref{SDelta}) is not unique. For instance, at one loop order $\varpi'=\varpi^{(1)}+ Z$, where $Z$ is an invariant
operator, $[Z,S_\alpha^{(0)}]=0$, also satisfies Eqs.~(\ref{SDelta}).

Closing this section we give the standard representation for the conformal operator,
\begin{align}
\mathcal{O}_j(x)=P_j(\partial_{z_1},\partial_{z_2})[\mathcal{O}(x;z_1,z_2)]\Big|_{z_1=z_2=0}\,.
\end{align}
The operator $\mathcal{O}_j$ is completely determined by
a polynomial   $P_j(z_1,z_2)$. It was known a long ago~\cite{Makeenko:1980bh} that at the leading order
$$
P_j(z_1,z_2)\sim (z_1+z_2)^j C_j^{(3/2)}\left(\frac{z_1-z_2}{z_1+z_2}\right),
$$
where $C_j^{(3/2)}$ is the Gegenbauer polynomial. Beyond the leading order one derives from Eqs.~(\ref{scalar-O})
and (\ref{scalar-exact})
\begin{align}\label{P-exact}
P_j(z_1,z_2)&=(z_1+z_2)^j\biggl\{p_j^{(\lambda)}(z_1,z_2)
\notag\\
&\quad -4\,u_*\lambda_s
\sum_{k=0,2\ldots}^{j} b_k^j\, p_{k}^{(\lambda)}(z_1,z_2)\biggr\}\,,
\end{align}
where $\lambda=2s_*-1/2$,
\begin{align*}
p_j^{(\lambda)}(z_1,z_2)= \frac{j!\Gamma(2\lambda)\Gamma(\lambda+\frac12)}{\Gamma(j+2\lambda)\Gamma(j+\lambda+\frac12)}
C_j^{(\lambda)}\left(\frac{z_1-z_2}{z_1+z_2}\right),
\end{align*}
and the expansion coefficients have the form
\begin{align}
&b_j^j=\frac{1}{(j+1)(j+2)}\left[S_{2j+2}-2S_{j+1}-\frac{1}{(j+1)(j+2)}\right],
\notag\\
&b_{k<j}^j=\frac{(2k+3)}{(j-k)(j+k+3)}\frac{(k+1)!}{(j+2)!(j+1)}\,.
\end{align}
The expression~(\ref{P-exact}) agrees with the expression for the conformal operator obtained in \cite{Mueller:1991gd,Belitsky:2007jp}.

\section{Loop diagrams}\label{app:D}
We will present an  answer for a diagram $D^{(a)}$ minus the counterterm $\Delta D^{(a)}$ in the form
\begin{align*}
x^2(n\bar n)\,(D^{(a)}-\Delta D^{(a)})=u_*^2 \cdot \varkappa_j T_j^{(a)}\vev{\mathcal{O}_j^{(n)}(x)\mathcal{O}_j^{(\bar n)}}_{0}\,.
\end{align*}
Here
$\vev{\mathcal{O}_j^{(n)}(x)\mathcal{O}_j^{(\bar n)}}_{0}$ is the tree level correlator.  
For the diagrams in Fig.~\ref{fig:c33A1} and ~\ref{fig:c33B1}  we obtain
\begin{align}\label{T5}
T_j^{(5,A)}&=\frac{\lambda_s\lambda_d}{12}\biggl[S_{2j+2}-S_{j+3}-S_{j+2}+\frac73\biggr]\,,
\notag\\
T_j^{(5,B)}&=\frac{\lambda_s\lambda_d}{(j+1)(j+2)}\biggl[S_{2j+2}-S_{j+3}+\frac{1}{j+1}\biggr],
\notag\\
T_j^{(5,C)}&=-
\frac{2\lambda_s\lambda_d}{(j+1) (j+2)}
\biggl[S_{2j+2}-S_{j+3}-\frac12{S_{j}}+1\biggr]\,,
\notag\\
T_j^{(5,D)}&=-\frac{\lambda_s\lambda_d}{(j+1)^2(j+2)^2}\,,
\end{align}
and
\begin{align}\label{T6}
T_j^{(6, A)}&=
 -\gamma_j^{(1)}
\biggl\{\gamma_j^{(1)}\left[S_{2j+2}-S_{j+3}-S_{j+2}+\frac53\right]
\notag\\[2mm]
&\quad
-\frac{2\lambda_s}{(j+1)(j+2)}
\left[S_{j+2}-\frac{(2j+1)(4j+7)}{3(j+1)(j+2)}\right]\biggr\}\,,
\notag\\[2mm]
T_j^{(6, B)} &=-\frac12  (\gamma_j^{(1)})^2\frac{j^2+3j+4}{(j+1)(j+5)}\,.
\end{align}

%


\end{document}